# Big Deformation in $^{17}$C

FAN Guang-Wei(樊广伟)[1,*]   CAI Xiao-Lu(蔡晓鹭)[2]   M. Fukuda[3]
HAN Ti-Fei(韩体飞)[1]   LI Xue-Chao(李学超)[1]   REN Zhong-Zhou(任中洲)[4]
XU Wang(徐望)[2]

1 (School Of Chemical Engineering, Anhui University of Science and Technology, Huainan, 232001, China)
2 (Shanghai Institute of Applied Physics, Chinese Academy of Sciences, Shanghai 201800, China)
3 (Department of Physics, Osaka University, Osaka 560-0043, Japan)
4 (Department of Physics, Nanjing University, Nanjing 210008, China)

**Abstract:** Reaction- and interaction cross sections of $^{17}$C on a carbon target have been reanalyzed by use of the modified Glauber model. The analysis with a deformed Woods-Saxon density/potential suggests a big deformation structure for $^{17}$C. The existence of a tail in the density distribution supports its possibility of being a one-neutron halo structure. Under a deformed core plus single-particle assumption, analysis shows a dominant $d$-wave of the valence neutron in $^{17}$C.

## 1 Introduction:

$^{17}$C, with small one-neutron separation energy $S_n = 0.729 \pm 0.018$ MeV and large two-neutron separation energy $S_{2n} = 4.979 \pm 0.018$ MeV [1], is an interesting candidate for a one-neutron halo nucleus. Since without the Coulomb barrier, the valence neutron separation energy could mostly confirm neutron-halo structure. $^{17}$C is a typical $psd$-shell nucleus, the valence neutron radial wave function exhibits configuration mixing of the $s$ and $d$-wave. If the valence neutron has a $d$-dominant configuration, the radial extension of the wave function will not be significant [2].

Early experimental studies suggested not a possible halo structure for $^{17}$C. The momentum distribution of the fragment $^{16}$C from $^{17}$C was found to be relatively broad [3-5]. The interaction cross section ($\sigma_I$) at 965 MeV/A did not show a significant enhancement to its neighbors [6]. These indicated that there was no halo-structure for $^{17}$C. However, subsequent experimental studies gave a conflicting result. The measurement of the reaction cross section ($\sigma_R$) by C. Wu et al. [7] for $^{17}$C on $^{12}$C at 79 MeV/A suggested that $^{17}$C was a one-neutron halo nucleus. Finally, they showed us the necessity of a long tail structure for $^{17}$C by use of the Glauber-type analysis.

This confliction reminds us whether there is a big deformation for $^{17}$C, since the deformation can also much contribute to $\sigma_R$ and $\sigma_I$ [8]. Besides, Shen Yao-song et al. [9] claimed the deformation for $^{17}$C by the calculation of the Deformed-Skyrme-Hartree-Fock model. These motivated us to reanalyze the experimental data of $^{17}$C. In this article, we will use the modified Glauber model to reanalyze the experimental data and finally extract the density distribution of $^{17}$C. With the result, we can address the confliction.

## 2 Formalism of the Modified Glauber Model

The optical limit Glauber model, given by Glauber R J [10], is a useful tool to connect $\sigma_R$ (and $\sigma_I$) with a nucleon density distribution, though the model underestimates the $\sigma_R$ at low energies because the multiple scattering effect and Fermi-motion are not taken into account. Therefore, we adopted the modified optical limit Glauber model (MOL), an improvement proposed by Abu-Ibrahim and Suzuki [11], and Takechi M et al. [12]. With this improved Glauber model, we reanalyzed the experimental $\sigma_R$ and $\sigma_I$, deduced the nucleon density distribution of $^{17}$C through a $\chi^2$-fitting procedure.

The MOL used in the present analysis was described in detail in Ref. [12], and formulated as follows. The $\sigma_R$ is given by

$$\sigma_R = 2\pi \int db\, b[1 - T(b)]C(E), \tag{1}$$

where $C(E)$ denotes the influence of the Coulomb force [13]. T(b) denotes the transmission probability at an impact parameter $b$. In the MOL, T(b) is expressed as

$$T(b)^{MOL} = \exp\{-\int ds\, \rho_z^P(s)(1 - \exp[\int dt\, \rho_z^T(t)\sigma_{NN} \times \Gamma(b+s-t)])\}\exp\{-\int dt\rho_z^T(t) \times (1 - \exp[\int ds\rho_z^P(s)\sigma_{NN}\Gamma(b+t-s)])\}, \tag{2}$$

where $\rho_z^P$ and $\rho_z^T$ are the z-integrated densities of the projectile and the target nuclei, respectively, $\sigma_{NN}$ is the nucleon-nucleon total cross section at kinetic energy, $\Gamma$ is the nucleon-nucleon profile function, and $s$, $t$ are the nucleon coordinates of the projectile and the target in the plane perpendicular to the beam axis. In the MOL, $\sigma_{NN}$ is corrected by the effective $\sigma_{NN}$, which is described as:

$$\sigma_{NN}^{eff} = \int_{-\infty}^{\infty} dP_{rel}\sigma_{NN}D(P_{rel}), \tag{3}$$

where D($P_{rel}$) is expressed with relative momentum between nucleons in the projectile and the target as

$$D(P_{rel}) = \frac{1}{\sqrt{2\pi(\langle P_P^2\rangle + \langle P_T^2\rangle)}} \times \exp\left[-\frac{(P_{rel}-P_{proj})^2}{2\sqrt{\langle P_P^2\rangle + \langle P_T^2\rangle}}\right]. \tag{4}$$

In this equation, $P_{proj}$ denotes the momentum of a nucleon with the same velocity as the projectile, $\langle P_P^2\rangle$ a mean square momentum of a nucleon in the projectile and $\langle P_T^2\rangle$ that in the target. For stable nuclei, we employed the averaged experimental value of 90 MeV/c as $\sqrt{\langle p_T^2\rangle}$. For $^{17}$C, the $\rho^n$ in Eq. (2) was divided into a core and one valence nucleon part. For the core part we used the experimental value of momentum width from the data for $^{16}$C (= 73 MeV/c), and for the valence part the data for $^{17}$C (= 61 MeV/c) [3].

## 3  Nuclear density distribution of $^{17}$C

Like Wu C et al. [7], the density function of $^{17}$C was divided into a core ($^{16}$C) and a valence neutron part, a spherical harmonic oscillator (HO) type function was used as the core shape, the Yukawa function and single particle model (SPM) density were used as the valence neutron shape.

**The HO type function**

$$\rho_c^i(r) = \rho_{c0}^i \times \left(1 + \frac{C-2}{3}\left(\frac{r}{b}\right)^2\right), \tag{5}$$

where $i$ denotes the proton or neutron and $C$ is the number of proton of neutron in the core. The $b$ is the core width parameter and $\rho_{c0}$ is the normalization factor. The same width was used for the proton- and neutron-core densities.

**The Yukawa function**

For protons

$$\rho^p(r) = \rho_c^p(r), \quad (6)$$

For neutrons

$$\rho^n(r) = \begin{cases} X \times \rho_c^n(r) & r \leq r_c \\ Y \times \frac{exp(-\lambda r)}{r^2} & r > r_c \end{cases} \quad (7)$$

where $r_c$ is the intersection point of the core and the tail part, $\lambda$ the tail slope, and $X$ and $Y$ the amplitude of the core and the tail part, respectively. In the $\chi^2$-fitting process, we assumed that $b$ (= 1.778 fm) was the same as that of $^{16}$C [14].

**Single particle model**

In SPM, the wave function of the valence neutron was calculated by solving the Schrödinger equation numerically, assuming the Woods-Saxon (WS) potential, the Coulomb barrier, and the centrifugal barrier. The nuclear part of the potential assumed is written as

$$V = \left(-V_0 + V_1(l \cdot s)\frac{r_{ls}^2}{r}\frac{d}{dr}\right)\left[1 + exp\left(\frac{r-R_c}{a}\right)\right]^{-1}, \quad (8)$$

where $a$ (= 0.70 fm) and $R_c$ (= $r_0 A^{1/3}$, $r_0$ = 1.22 fm) are the diffuseness parameter and radius of the WS potential [15]. The depth of this potential was adjusted to reproduce the experimental binding energy of the valence neutron. $^{17}$C is a typical *psd*-shell nucleus, the valence neutron radial wave function exhibits configuration mixing of the *s* and *d*-wave. We assumed that the neutron density of $^{17}$C consisted of a $^{16}$C core plus a neutron with a mixing of the *s*-wave and the *d*-wave. In this case, $S_n$ is a free parameter and is assumed to be in range from 0.729 MeV to 0.729 + 1.766 MeV (1.776 MeV is the excitation energy). We searched for the minimum $\chi^2$-fit between the low- and high energy data by varying the ratio of the *s*- and *d*-wave. A proportion of 73 ±24% for the *d*-wave was found when the $\chi^2$ reached the minimum.

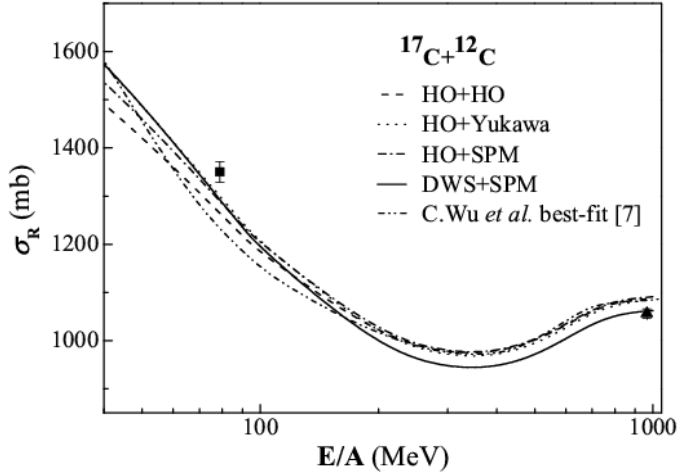

**Fig. 1.** The $\sigma_R$ data for $^{17}$C as a function of beam energy. The experimental data of closed square was taken from Ref. [7] and closed triangle was taken from Ref. [6].

Fig. 1 shows results of the analysis with HO + HO, HO + Yukawa and HO + SPM type functional shape. The minimum $\chi^2$ is 10.2 obtained by the analysis with HO + Yukawa function. In Fig. 1, large under- and overestimation of the calculation are found with the analysis at low- and high energy, which means these kinds of density distributions are not sufficient to describe the density distribution of $^{17}$C. However, it also shows us that the results of the analysis with HO + Yukawa and HO + SPM are a little better than that of HO + HO, especially, in the low energies, which means that a tail structure is necessary to describe the density distribution of $^{17}$C, since the

$\sigma_R$ is more sensitive to the surface density part at low energies. So we try to test the deformed core plus tail to describe the density of $^{17}$C. In order to keep the consistency of the core, the deformed WS (DWS) distribution was chosen to describe the density of the core. It is expressed as

$$\rho(r) = \frac{\rho_0}{1+exp\left(\frac{r-R(\theta)}{a}\right)}, \qquad (9)$$

where $R(\theta) = R_c\left(1 + \beta Y_{20}(\theta)\right)$, $R_c$, $a$ and $\beta$ were chosen to reproduce the quadrupole momentum (= 46.05 fm$^2$) given by the Deformed-Skyrme-Hartree-Fock Calculation of $^{17}$C [9]. The WS potential in the SPM was corrected by $R(\theta)$ too. A proportion of 70 ± 21% for the $d$-wave was found when $\chi^2$ reached the minimum 6.5. The $d$-wave dominant was consistent with the calculation of Maddalena V et al. [16, 17] and Datta Pramanik U et al. [18]. The density distribution extracted was shown in Fig. 2. The error of the density for $^{17}$C was obtained by total $\chi^2$ +1 (= 7.5) method.

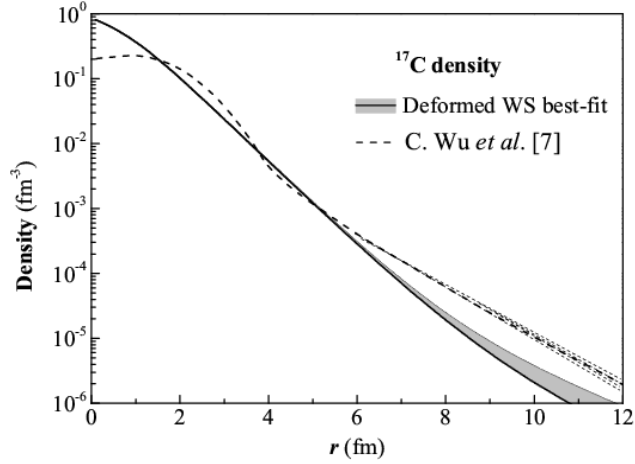

Fig. 2, Density distribution of $^{17}$C deduced by modified Glauber with deformed WS core plus SPM type functional shape. The center of mass effect was taken into account.

The result of the analysis with DWS density is shown in Fig. 1. It exhibits us that the analysis with DWS density is much better than that with spherical core plus tail density, which indicates the necessity of the deformation for $^{17}$C. Fig. 2 shows the density distribution of $^{17}$C. It shows us that $^{17}$C has a tail structure, though a $d$-wave dominant configuration hinders the radial extension of the wave function. Although, the definition of halo structure is still ambiguity, we can conclude that $^{17}$C is a mostly halo-like nucleus. The deformation may explain the broad momentum distribution of the fragment $^{16}$C from $^{17}$C. In order to investigate the reason for the broad momentum distribution, more experimental and theoretical works are needed.

## 4  Summary

We have reanalyzed reaction- and interaction cross sections of $^{17}$C on a carbon target using the well tested modified Glauber model. The results of the analysis show us that $^{17}$C has a big deformation and a tail structure. Based on the assumption of a deformed core plus a valence neutron, it is found that the valence neutron of $^{17}$C is mostly in the $d$-orbital.

### Acknowledgments

The authors want to thank Prof. Fang De-Qing, Chen Jin-Gen (In SINAP) and doctor D. Nishimura (In Tokyo University) for their helps in this subject. We would like to acknowledge the financial support provided by Anhui University of Science and Technology (11130).